\documentclass[twocolumn]{aastex61}

\newcommand\aastex{AAS\TeX}

\shorttitle{Average Spectral Properties of Type Ia Supernova Host Galaxies}
\shortauthors{Uddin, Mould, and Wang}
\begin{document}

\title{Average Spectral Properties of Type Ia Supernova Host Galaxies}

\correspondingauthor{Syed A  Uddin}
\email{saushuvo@gmail.com}

\author{Syed A Uddin}
\affiliation{Purple Mountain Observatory, Chinese Academy of Sciences, Nanjing, Jiangshu, China}
\affiliation{ARC Centre of Excellence for All-Sky Astrophysics (CAASTRO), Australia }

\author{Jeremy Mould}
\affiliation{ARC Centre of Excellence for All-Sky Astrophysics (CAASTRO), Australia }
\affiliation{Centre for Astrophysics and Supercomputing, Swinburne University of Technology, Melbourne, VIC, Australia}

\author{Lifan Wang}
\affiliation{Purple Mountain Observatory, Chinese Academy of Sciences, Nanjing, Jiangshu, China}
\affiliation{Department of Physics and Astronomy, Texas A$\&$M University, College Station, TX, USA }

%\collaboration{(AAS Journals Data Scientists collaboration)}

%\author{Butler Burton}
%\affiliation{National Radio Astronomy Observatory}
%\affiliation{AAS Journals Associate Editor-in-Chief}
%\nocollaboration
%
%\author{Amy Hendrickson}
%\altaffiliation{Creator of AASTeX v6.1}
%\affiliation{TeXnology Inc.}
%\collaboration{(LaTeX collaboration)}
%
%\author{Julie Steffen}
%\affiliation{AAS Director of Publishing}
%\affiliation{American Astronomical Society \\
%2000 Florida Ave., NW, Suite 300 \\
%Washington, DC 20009-1231, USA}
%
%\author{Jeff Lewandowski}
%\affiliation{IOP Senior Publisher for the AAS Journals}
%\affiliation{IOP Publishing, Washington, DC 20005}

%% Note that the \and command from previous versions of AASTeX is now
%% depreciated in this version as it is no longer necessary. AASTeX 
%% automatically takes care of all commas and "and"s between authors names.

%% AASTeX 6.1 has the new \collaboration and \nocollaboration commands to
%% provide the collaboration status of a group of authors. These commands 
%% can be used either before or after the list of corresponding authors. The
%% argument for \collaboration is the collaboration identifier. Authors are
%% encouraged to surround collaboration identifiers with ()s. The 
%% \nocollaboration command takes no argument and exists to indicate that
%% the nearby authors are not part of surrounding collaborations.

%% Mark off the abstract in the ``abstract'' environment. 
\begin{abstract}

We construct average spectra of host galaxies of slower, faster, bluer, and redder Type Ia Supernovae (SNe Ia) from the SDSS-II supernova survey. The average spectrum of slower declining (broader light-curve width or higher stretch) SN Ia hosts shows stronger emission lines compared to the average spectrum of faster declining (narrower light-curve width or lower stretch) SN Ia hosts. Using pPXF, we find that hosts of slower declining SNe Ia have metallicities that are, on average, 0.24 dex lower than average metallicities of faster declining SN Ia hosts. Similarly, redder SN Ia hosts have slightly higher metallicities than bluer SN Ia hosts. Lick index analysis of metallic lines and Balmer lines show that faster declining SN Ia hosts  have relatively higher metal content and have relatively older stellar populations compared with slower declining SN Ia hosts. We calculate average $\rm H_{\alpha}$ Star Formation Rate (SFR), stellar mass, and the specific-SFR (sSFR) of host galaxies in these subgroups of SNe Ia. We find that slower declining SN Ia hosts have significantly higher ($>5\sigma$) sSFR than faster declining SN Ia hosts. A Kolmogorov-Smirnov test shows that these two types of hosts originate from different parent distributions. Our results, when compared with the models of  \cite{childress14}, indicate that slower declining SNe Ia, being hosted in actively star-forming galaxies, are young (prompt) SNe Ia, originating from similar progenitor age groups. 

\end{abstract}
%% Keywords should appear after the \end{abstract} command. 
%% See the online documentation for the full list of available subject
%% keywords and the rules for their use.

%% From the front matter, we move on to the body of the paper.
%% Sections are demarcated by \section and \subsection, respectively.
%% Observe the use of the LaTeX \label
%% command after the \subsection to give a symbolic KEY to the
%% subsection for cross-referencing in a \ref command.
%% You can use LaTeX's \ref and \label commands to keep track of
%% cross-references to sections, equations, tables, and figures.
%% That way, if you change the order of any elements, LaTeX will
%% automatically renumber them.

%% We recommend that authors also use the natbib \citep
%% and \citet commands to identify citations.  The citations are
%% tied to the reference list via symbolic KEYs. The KEY corresponds
%% to the KEY in the \bibitem in the reference list below. 

%%%%%%%%%%%%%%%%%%%%

\section{Introduction}

Type Ia Supernovae (SNe Ia) are currently the most useful distance indicators across the wider history of cosmic expansion. Empirical correlations between the properties of SNe Ia have made them standardizable candles \citep{phillips93}. This eventually lead to the surprising discovery that the expansion of the universe is accelerating (\citealt{perlmutter97}, \citealt{riess98}). The unknown force responsible for this acceleration, commonly known as dark energy, is poorly understood. Reducing uncertainties in determining cosmological constraints is the key to shed light on the nature of dark energy (\citealt{scolnic17}; \citealt{betoule14}). Understanding the properties of SNe Ia and their astrophysical origin is an avenue to achieve this goal.

%The popular explanation for dark energy is that it is the cosmological constant (\citealt{weinberg89}) and current data tends to support this explanation. But this is fundamentally different than the known physics of our universe. On the other hand, the time varying nature of dark energy is hard to detect, let alone detecting if the gravity is different at cosmological scales. 

SNe Ia seem to show diversity in their light curve properties. Some are bluer while some are redder. Some have faster declining light curves, while some have slower declining light-curves. Why these diversities exist, is not clear yet. Theory suggests that SN Ia explosions are powered by radioactive decay of $\rm ^{56}Ni$.  More  $\rm ^{56}Ni$ produces more luminous, slower declining SNe Ia. While directly probing SN Ia progenitors is challenging, we can study the global stellar populations of SNe Ia as a proxy for the progenitors. For example, \cite{timmes03} suggest a relationship between  $\rm ^{56}Ni$ and host metallicity. If this prediction is true then we expect fainter, faster declining SNe Ia to be associated with massive, metal-rich, and therefore, lower star-forming galaxies and vice versa. \cite{neill09} have found that this trend is consistent with observed data.  

A number of studies have been carried out on SN Ia - host correlations using both global and local host properties (e.g., \citealt{uddin17}; \citealt{pan14}; \citealt{gupta11}; \citealt{lampeitl10a}; \citealt{sullivan10}; \citealt{rigault13}; \citealt{campbell16}; \citealt{wolf16}; \citealt{johansson13}; \citealt{hill16}; \citealt{ramon17} etc). Some of these studies use host photometry, while others use spectra of host galaxies. The general nature of these studies are two-fold: (1) SNe Ia are grouped according to their host galaxy properties and thereafter mean SN Ia properties are computed;  (2) linear regression lines are drawn between pairs of SN Ia-host property groups and the slopes are computed. From these studies it is established that SNe Ia, after standardization, are more luminous\footnote{By more luminous we mean relatively lower Hubble residual, which is defined as, $\Delta \mu = \mu_{obs} - \mu_{mod}$, the difference between the observed and the model distance modulus.} and faster declining in massive (metal-rich), older hosts and hosts with lower specific star-formation rates (sSFR). 

Why the above discussed correlations exist, is not clear yet. \cite{childress14} have made theoretical investigations on the ages of SNe Ia. Their study shows that young (prompt) SNe Ia are found in young, less massive, and actively star-forming galaxies. They originate from similar progenitor age groups and should make a uniform sample for cosmological analysis. On the other hand, old (tardy) SNe Ia are found in older galaxies with relatively lower star-formation activities. These SNe Ia originate from different progenitor age groups. It is also difficult to clearly understand progenitor channels of SNe Ia. Two possible scenarios are proposed: single degenerate and double degenerate. A recent analysis from  \cite{heringer17} favors a double degenerate case, yet other options are possible.

A way to understand the diversity in SN Ia properties is  to study average spectral signatures and average properties of their host galaxies grouped according to the properties of SNe Ia, such as color and light-curve width. The advantage of such analysis is that it will give a clear picture of the properties of stellar populations of SNe Ia with different properties. \cite{brandt10} has constructed average spectra of faster and slower declining SNe Ia hosts  from 101 SDSS-II SN Ia sample and showed that the average spectrum of slower declining hosts is dominated by stronger emission lines, indicating ongoing star-formation. Using a sample of 84 SNe Ia, \cite{johansson13} split SN Ia host spectra into four bins in stretch and found that as we go from lower to higher stretch, emission line strengths of the host galaxies become increasingly prominent.

While both above mentioned studies make qualitative statements, no quantitative results were shown. Their sample sizes were small and moreover they have only studied average spectra in stretch bins, not in SN Ia color bins. In this paper, we extend such study with 311 SNe Ia from SDSS-II -- a three-fold increase of sample size. Moreover, we quantify average emission line properties and derive average host properties, such as SFR and sSFR, after splitting SNe Ia according to their stretch and color.

In Section \ref{data} we describe our sample, in Section \ref{ana} we present our analysis of average spectra, spectral properties, and properties of hosts, and we summarise in Section \ref{sum}.

\section{Data}\label{data}
\begin{figure*}
\includegraphics[width=\columnwidth]{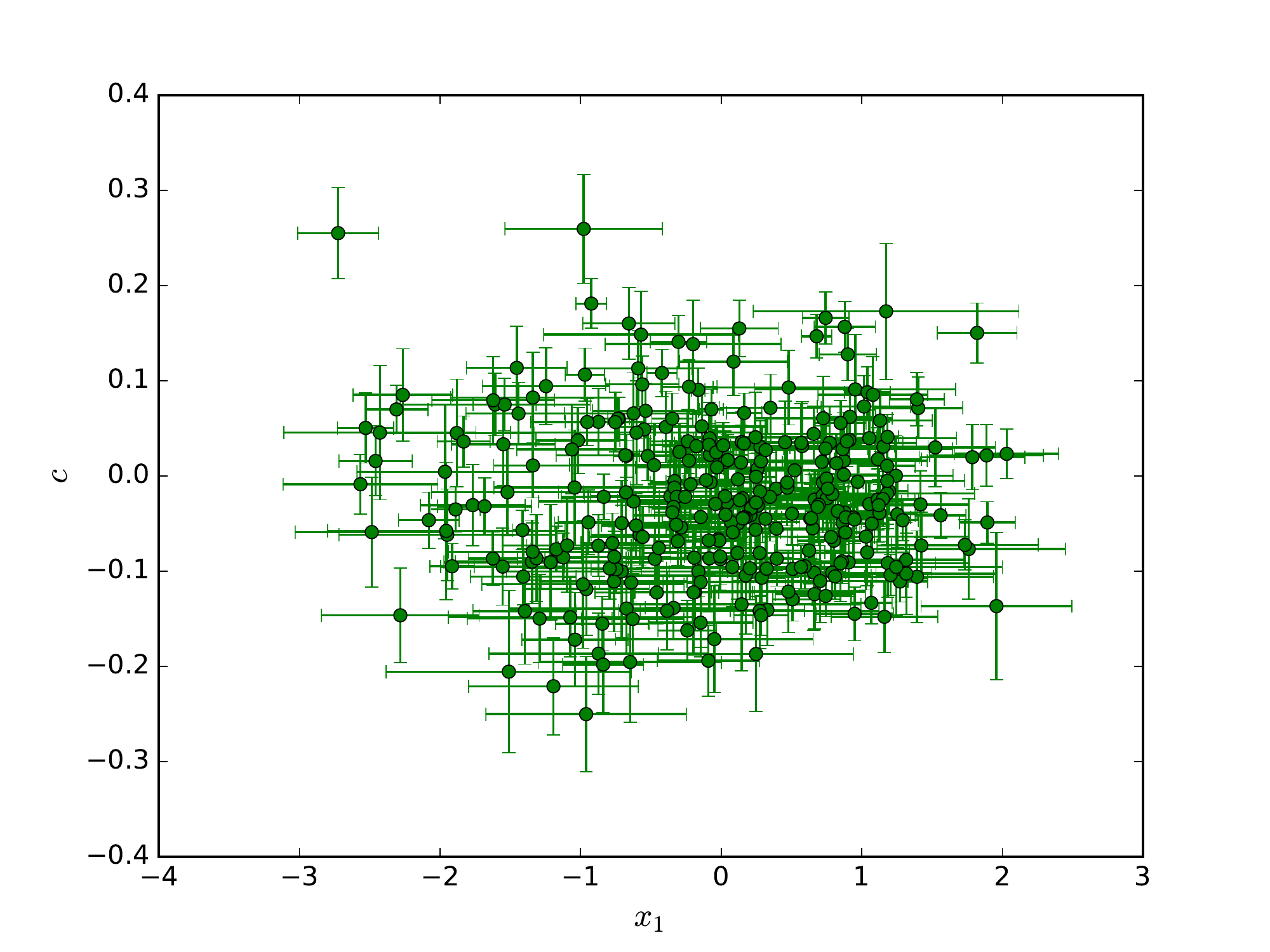} 
\includegraphics[width=\columnwidth]{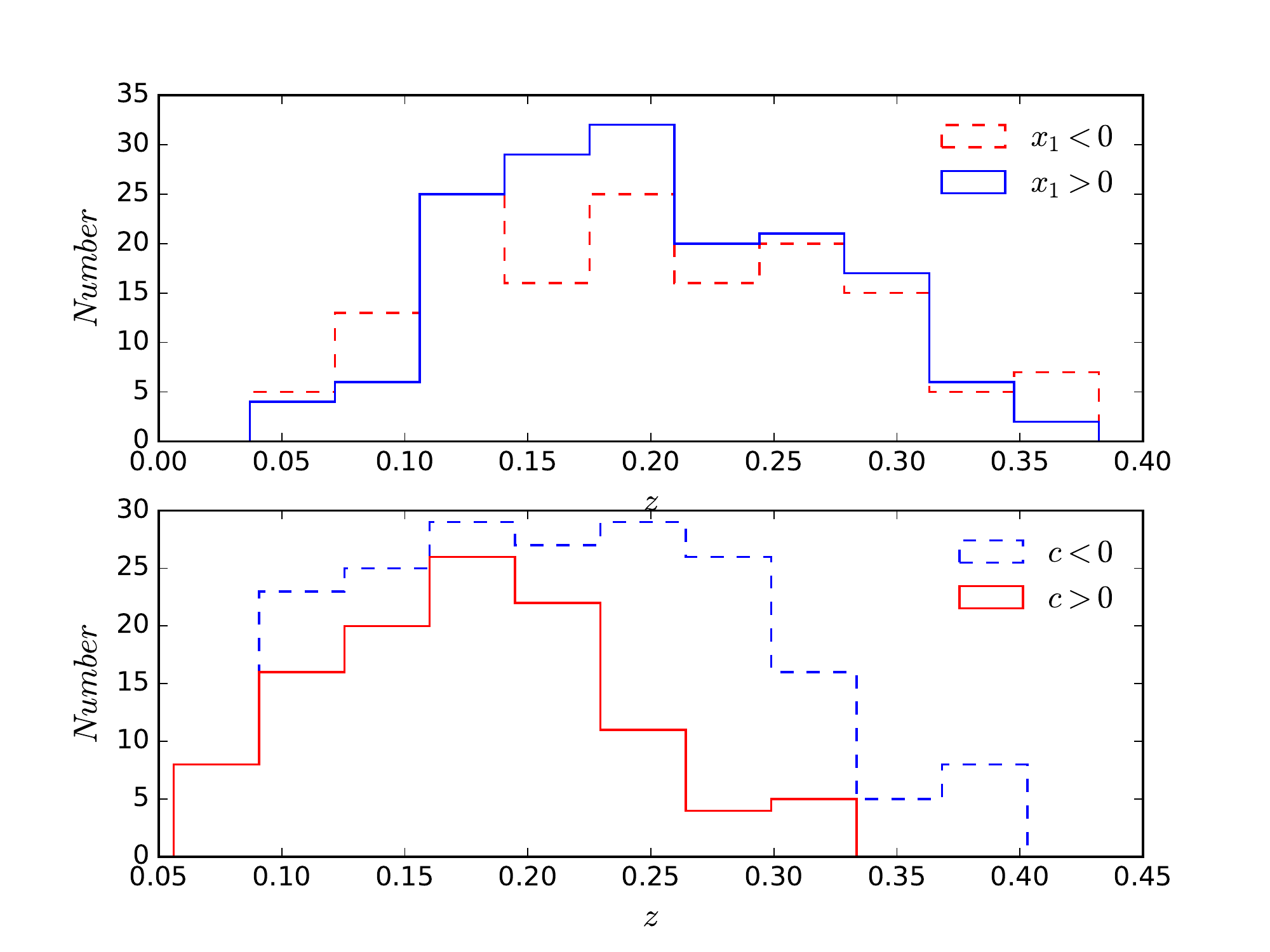} 
 \caption{SN Ia sample used in this work. Left: distributions in $x_1-c$ plane. Right: redshift distributions of SNe Ia. The top panel shows redshift distributions in two stretch subgroups and the bottom panel shows distributions in two color groups. We find an excess of bluer SNe Ia at higher redshifts due to Malmquist bias.}
\label{fig:data}
\end{figure*}

We use a sample of 311 SNe Ia from the Sloan Digital Sky Survey-II (SDSS-II; see \citealt{frieman08} for SN survey). In phase two of the survey SDSS-II scanned a part of the sky near the celestial equator, known as the Stripe 82 in order to discover supernovae. Between Sept 2005 - Nov 2007, SDSS-II has discovered $\sim$ 500 spectroscopically confirmed SNe Ia in the range $0.05<z<0.4$. The final data release of the SDSS-II supernova program is presented in \cite{sako14}. We have taken SN Ia properties such as redshift ($z$), stretch ($x_1$), and color ($c$) from the Joint Light curve Analysis (JLA; \citealt{betoule14}).  JLA uses SALT (\citealt{guy10}) to fit SN Ia light curves after calibrating with the CALSPEC standards. We show our sample in Fig.~\ref{fig:data}. The left panel shows the distribution of SNe in the stretch-color ($x_1- c$) plane. The right two panels show redshift distributions of SNe Ia in stretch and color subgroups. Subgroups are defined in Table \ref{group}. We see in color distributions that there are more bluer SNe Ia at higher redshifts than the redder ones. This is due to a Malmquist bias effect. The distributions in stretch subgroups seem to be similar. 

\begin{figure}
\begin{center}
\includegraphics[width=\columnwidth]{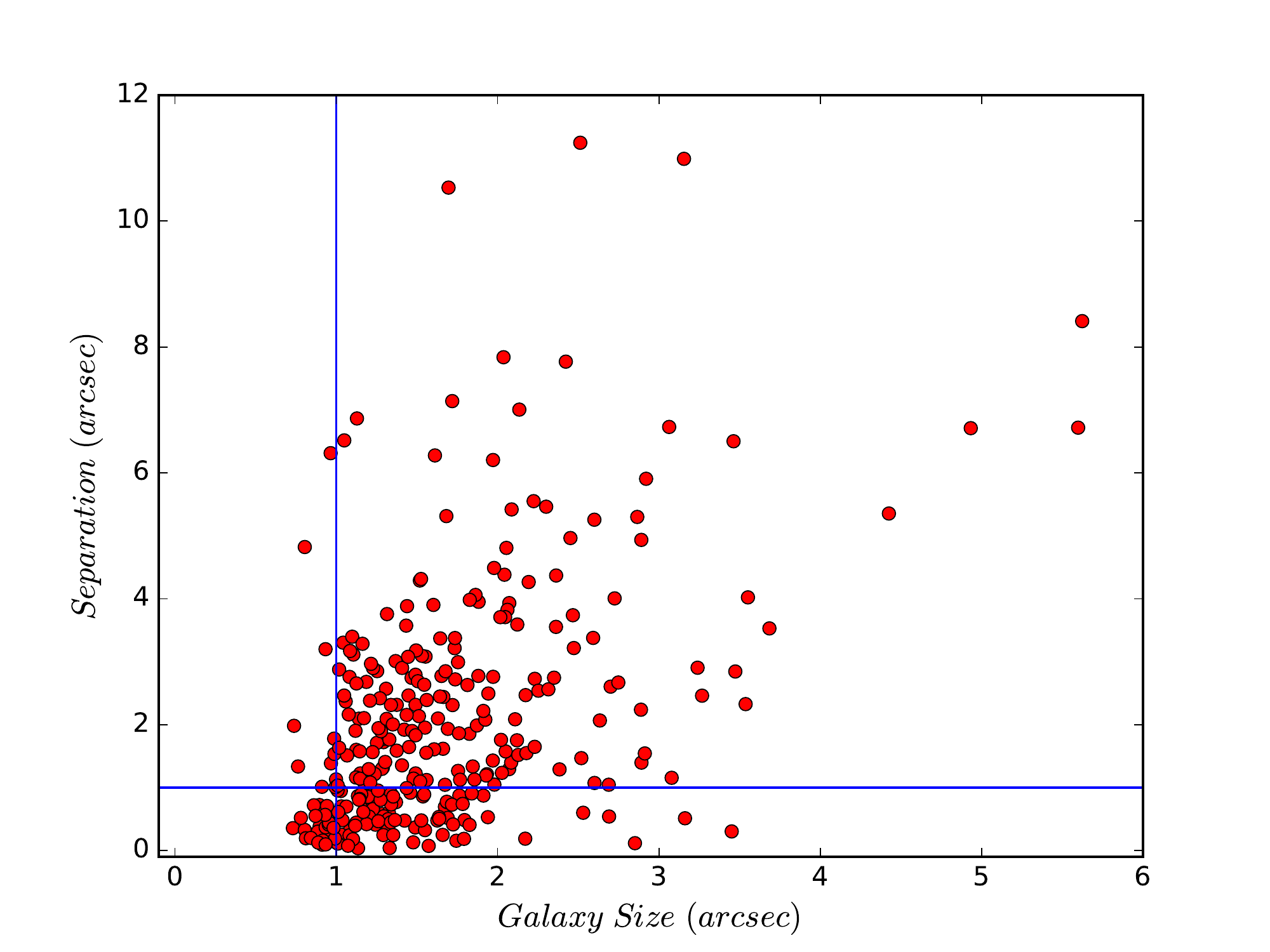}
\caption{Galaxy size and separation between SNe Ia and hosts. Here galaxy sizes are half-light radii obtained from SExtractor FLUX$\_$RADIUS parameter and by setting PHOT$\_$FLUXFRAC=0.5 in defining FLUX$\_$RADIUS. As noted in the text, $38\%$ of the time host galaxies fill the BOSS fibers of 1 arcsec radius. Also, just over a third of the time, separations between the SN Ia and host are within 1 arcsec.}
\label{size}
\end{center}
\end{figure}

The source of SN Ia host spectra is the Baryon Oscillation Spectroscopic Survey (BOSS; \citealt{dawson13}) - a survey conducted between fall 2009 to spring 2014. The spectrograph of BOSS has 1000 fibers each of diameter 2 arcsec. The resolution is $R\sim 2000$ and the wavelength coverage is 360-1000nm. Our goal is to study global properties of host galaxies, not the local properties at the explosion sites of SNe Ia. For the latter, integral field spectroscopy is promising. Fig.~\ref{size} shows that some host galaxies $(38\%)$ fall within 1 arcsec radius of the fiber. Also, $35\%$  of the time, separations between SNe Ia and host nuclei are within 1 arcsec.

 We gather FITS spectra from SDSS data release 12 (DR12) using a web based query located at the Science Archive Server\footnote{http://data.sdss3.org/}. The second extension of each FITS file contains a coadded calibrated spectrum of a given combination of Plate-MJD-Fiber, which is a unique combination for a given object. The fourth extension contains integrated fluxes and equivalent widths of various spectral lines.

\section{Analysis}\label{ana}

Our aim of this study is to compare average spectral properties and stellar populations of SN Ia hosts by splitting them according to SN Ia properties. We define various SN Ia subgroups in Table \ref{group} and split host galaxies according to these criteria. We first combine host spectra in different subgroups to create average host spectra and compare average spectra between subgroups. We then calculate average luminosities of various emission lines. Thereafter, we calculate SFR for these hosts, and using stellar mass from other work we calculate sSFRS. We present our analysis below.

\begin{table}
\caption{SN Ia subgroups used in this paper. }
\begin{center}
\begin{tabular}{lcc}
\hline
Subgroup & Criteria & Number\\
\hline
Faster & $x_1<0$ & 116\\
Slower & $x_1>0$ & 121\\
Bluer & $c<0$ & 145\\
Redder & $c>0$ & 92\\
\hline
\end{tabular}
\end{center}
\label{group}
\end{table}%

\subsection{Average Spectra}

\begin{figure*}
	\includegraphics[width=\textwidth]{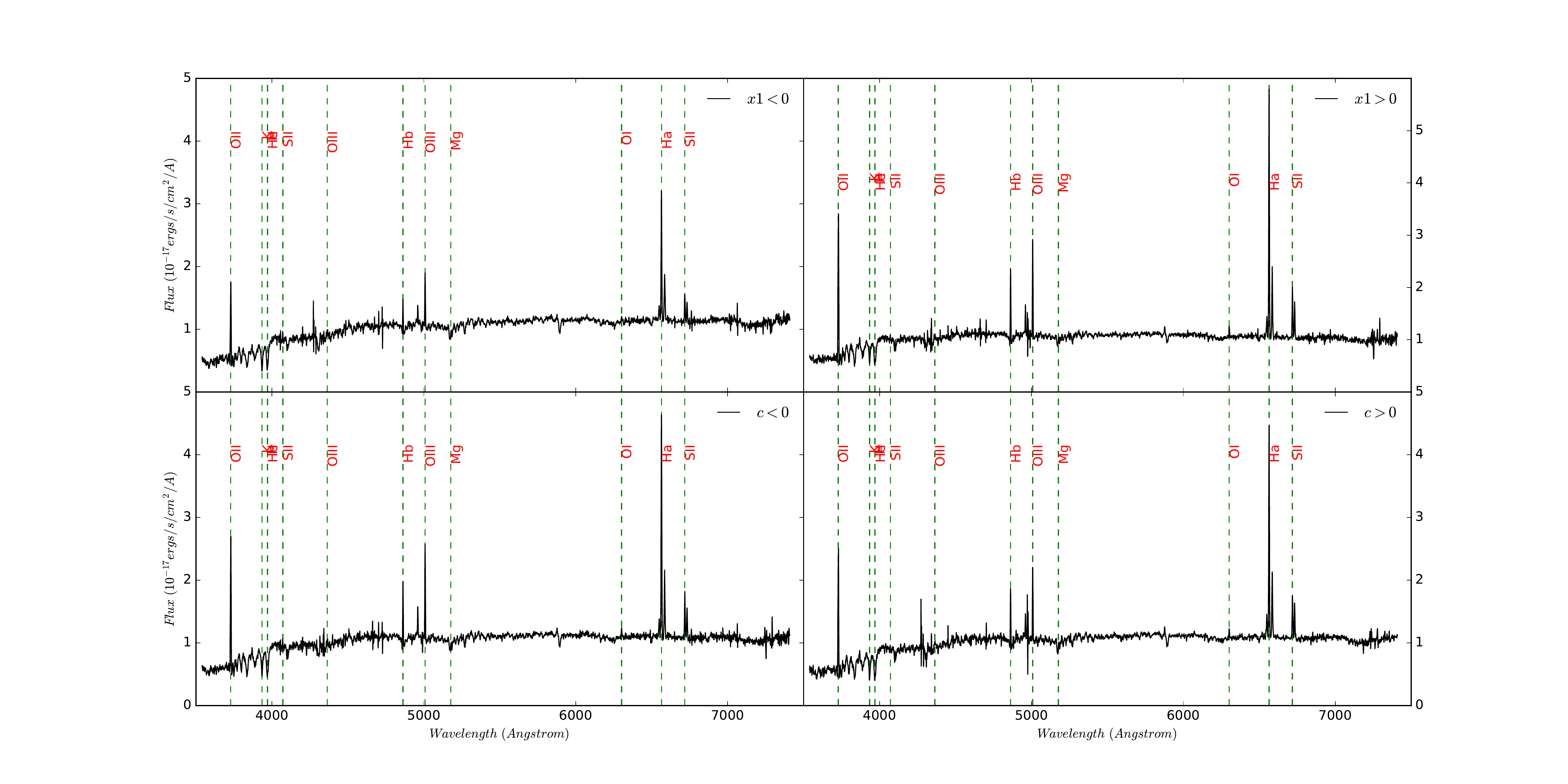}
    \caption{Top: Average spectra of SN Ia hosts split according to SN Ia stretch ($x_1$). Hosts in $x_1>0$ SNe Ia have relatively stronger emission lines compared to $x_1<0$ hosts.  Bottom: Average spectra of SN Ia hosts split according to SN Ia color ($c$). We see no significant difference in emission line strengths between two subgroups.} 
    \label{fig:x1c_ind}
\end{figure*}

In order to study average spectra of SN Ia hosts in each group, we first split them according to SN Ia properties and coadd. Before coadding we interpolate each spectrum on to the rest frame wavelengths. We then coadd observed spectra with no correction for distance effects, which gives higher weight to galaxies of greater apparent brightness.  Average spectra of SN Ia hosts in each subgroup are shown in Fig.~\ref{fig:x1c_ind}. 

It is easy to see from the top two panels of Fig.~\ref{fig:x1c_ind} that, on average, slower ($x_1>0$) SN Ia hosts have stronger emission lines compared to faster ($x_1<0$) SN Ia hosts, indicating stronger SFRs in them. This result confirms earlier findings of \cite{brandt10} and \cite{johansson13}. We do not see any significant difference in the average spectra in color subgroups as shown in the bottom two panels of Fig.~\ref{fig:x1c_ind}. We use penalized PiXel Fitting (pPXF; \citealt{cappellari04}) method to fit stellar continuum and emission lines in these average spectra. Ages and metallicities derived from pPXF fitting on these average spectra are given in Table \ref{tab:em2}. Slower declining SN Ia hosts have relatively lower metallicity (0.24 dex lower) than faster declining SN Ia hosts. We find them slightly younger than their counterparts. 

We  also calculate Lick indices (\citealt{worthey94}) to compare ages and metallicities that are shown in Table \ref{tab:em3}. Comparing metallic lines ($\rm Fe \ 4383$ \ and \ $\rm Mg_2$), we find that hosts of faster declining SNe Ia have higher equivalent widths compared with hosts of slower declining SNe Ia. This proves that faster declining SN Ia hosts are relatively more metal rich and have relatively older stellar populations than their counterparts. In terms of Balmer lines ($\rm \ H_{\beta} \ and \ H_{\delta}$), they are stronger in slower declining SN Ia hosts, indicating relatively younger and metal poor stellar populations. From both pPXF fitting and Lick index measurements, we find that the redder SN Ia hosts have higher metallicities than bluer SN Ia hosts. This result is consistent with \cite{childress13}, where they find that SNe Ia  are redder when they explode in higher metallicity hosts.

%Using photometry of the same sample, \cite{uddinphd} found that the average age of faster declining SN Ia hosts is 9.20 Gyr and the average age of slower declining SN Ia hosts is 8.92 Gyr.

% Example table
\begin{table}
	\centering
	\caption{Ages and metallicities of SN Ia hosts in various subgroups derived from average spectra using pPXF.}
	\label{tab:em2}
	\begin{tabular}{lcccc} % four columns, alignment for each
		\hline
		& $x_1<0$ & $x_1>0$ & $c<0$  & $c>0$\\

		\hline
		$\rm Log \ Age (Gyr)$                         & 0.875 & 0.902 &   0.926  & 0.880\\
		$\rm Metallicity \ [M/H]$                           & -0.163 & -0.403 &  -0.304  &  -0.238\\
		
		\hline
	\end{tabular}
\end{table}

We calculate average luminosities of various emission lines in various subgroups and show them in Table \ref{tab:em1}. We find that hosts of slower declining SNe Ia on average are more luminous than hosts of faster declining SNe Ia. We also find that hosts of bluer SNe Ia are also more luminous than those of redder SNe Ia.

\begin{table}
	\centering
	\caption{Lick indices measured from average spectra of various SN Ia subgroups. Negative signs in $H_{\beta}$ values indicate that the line is in emission. Errors in measurements are about $1\%$.}
	\label{tab:em3}
	\begin{tabular}{lcccc} % four columns, alignment for each
		\hline
		Equivalent Widths& $x_1<0$ & $x_1>0$ & $c<0$  & $c>0$\\

		\hline
		$\rm Fe \ 4383  \ (\r{A})$                       & 3.109 & 1.987 &   2.061  & 3.206\\
		$\rm H_{\beta} \ (\r{A})$                           & -0.090 & -2.437 &  -1.329  & -1.339\\
		$\rm H_{\delta}  \ (\r{A})$                           & 2.464 & 2.923 &  2.821  &  2.560\\
		$\rm Mg_2  \ (mag)$                           & 0.155 & 0.113 &  0.131  &  0.136\\

		\hline
	\end{tabular}
\end{table}

\begin{figure*}
	\includegraphics[width=\textwidth]{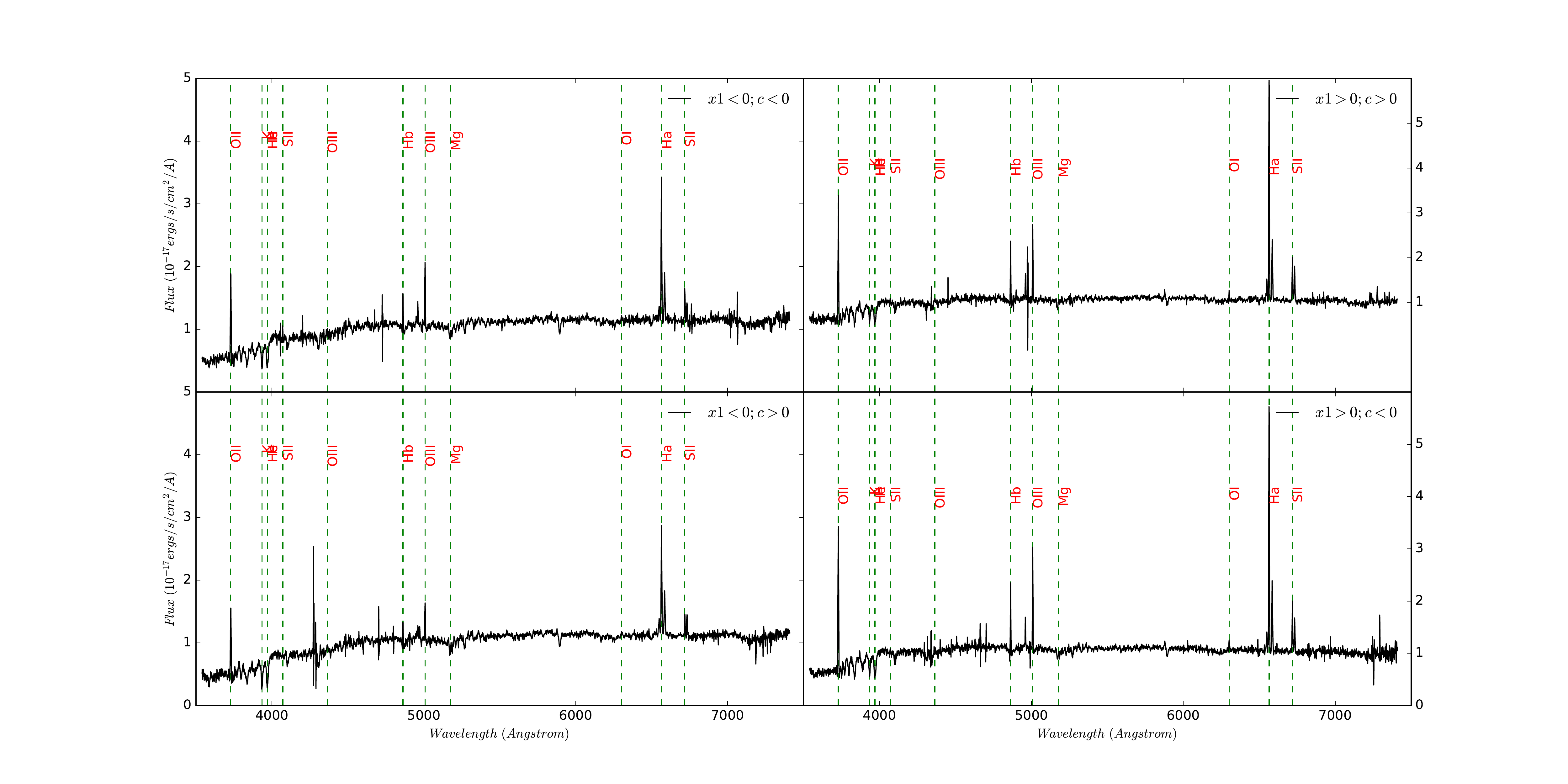}
    \caption{Average spectra of SN Ia hosts split according to four quadrants in $x_1-c$ plane. Here the observations are similar to what we see in Fig.~\ref{fig:x1c_ind}, but now with fewer host spectra in each stack. }
    \label{fig:x1c}
\end{figure*}

We further divide SN Ia hosts into four quadrants in the $x_1-c$ plane. Average spectra of them are shown in Fig.~\ref{fig:x1c}. For faster SNe Ia, bluer and redder hosts seem to differ in the strength of the emission lines.  Thus SFR (or more precisely, sSFR) are different. Redder ones have weaker emission. We expect that the redder ones would have more dust arising from more recent star formation, but we see the opposite here. Probably, there are more metal-rich and/or older.

For slower SNe Ia, there appears to be effectively no difference in host spectra between bluer and redder SNe Ia. If true, this would indicate that there is no real progenitor-related mechanism causing the color differences in these slower SNe Ia.  Thus there is probably something driving intrinsic color which may be completely stochastic.  It is a promising result for cosmology because it means we don't have to worry about intrinsic color evolution with redshift.

% Example table
\begin{table}
	\centering
	\caption{Luminosities of various emission lines in subgroups. Emission line luminosities are relatively higher for hosts of slower declining SNe Ia when compared with the hosts in faster declining  ones. Similarly, bluer SN Ia hosts are relatively more luminous than redder SN Ia hosts. }
	\label{tab:em1}
	\begin{tabular}{lcccc} % four columns, alignment for each
		\hline
		Emission Lines & $x_1<0$ & $x_1>0$ & $c<0$  & $c>0$\\
		($\rm \times10^{40}  \ Ergs/s$) &&&&\\
		\hline
		H$_\alpha$                         & 6.77 & 10.91 &   10.25  & 7.09\\
		H$_\beta$                           & 1.45 & 2.66 &  2.67  &  1.40\\
		H$_\gamma$                      & 0.75 & 1.18 & 1.26    & 0.64\\
		H$_\delta$                          &  1.12 & 2.45 &  3.17    & 0.61\\
		$\mathrm{[OII]}$3729        & 1.34 & 2.21 &   2.27  & 1.27\\
		$\mathrm{[NII]}$6583      & 3.73 & 4.00 &  3.89   & 3.84 \\
		\hline
	\end{tabular}
\end{table}

\subsection{Comparing Average Properties}
After studying average spectra we now study several properties of SN Ia hosts in various subgroups. First we calculate SFR. $\rm H_{\alpha}$ is considered to be the best choice for calculating SFR because of its intrinsic strength. It is also suitably located in the redder part of the spectrum that avoids significant dust extinction. We obtain integrated flux of $\rm H_{\alpha}$ emission line for each galaxy that was used in the coadds. We then calculate luminosity of  $\rm H_{\alpha}$ in $\Lambda$CDM cosmological framework. To calculate SFR we use an empirical formulation given by \cite{kennicutt98}:

\begin{equation}
\rm SFR =7.9\times 10^{-42}  \times L(H_{\alpha}) M_{\odot}  \ \rm yr^{-1}
\end{equation}

\begin{figure*}
	\includegraphics[width=\columnwidth]{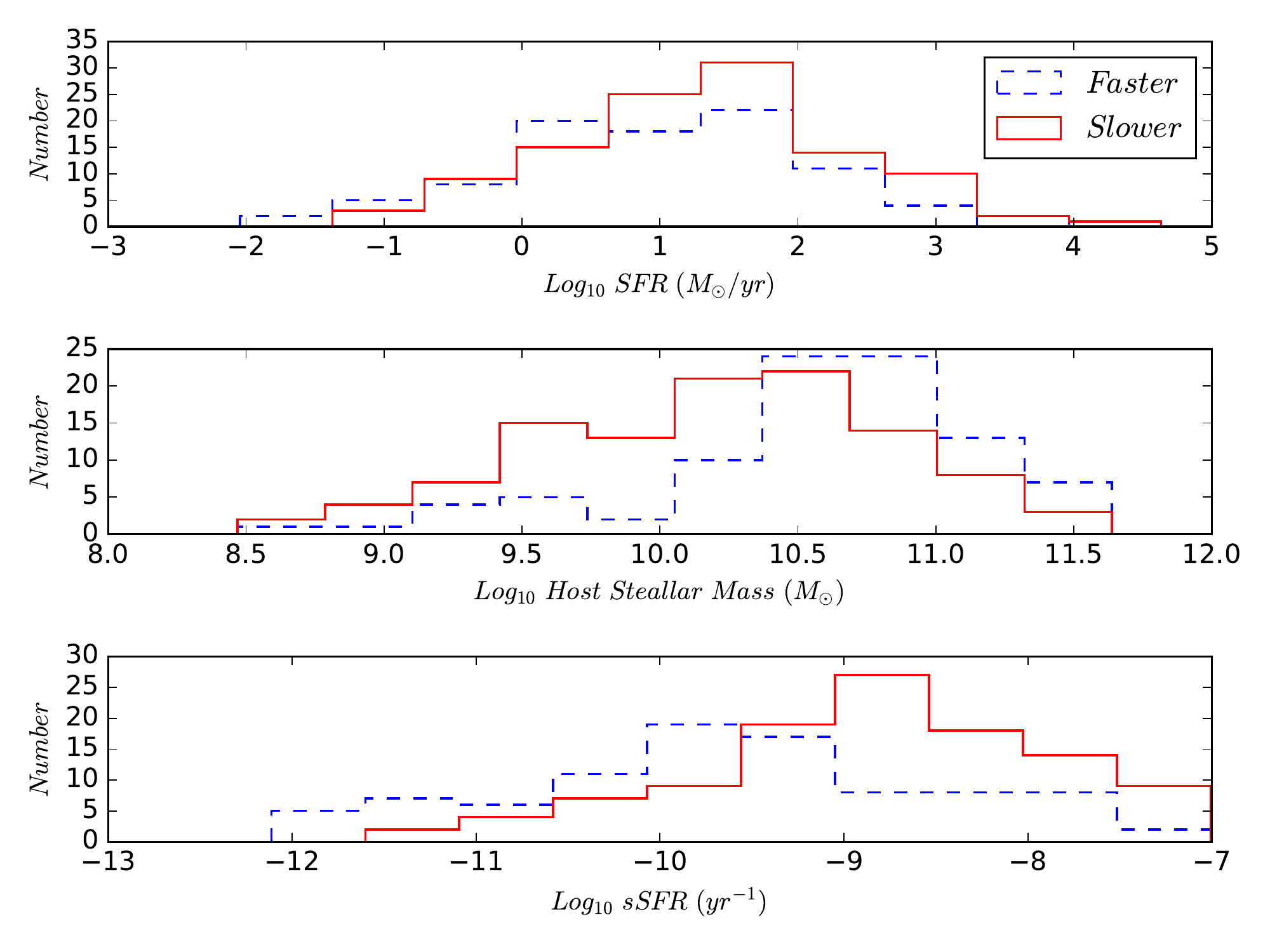}
	\includegraphics[width=\columnwidth]{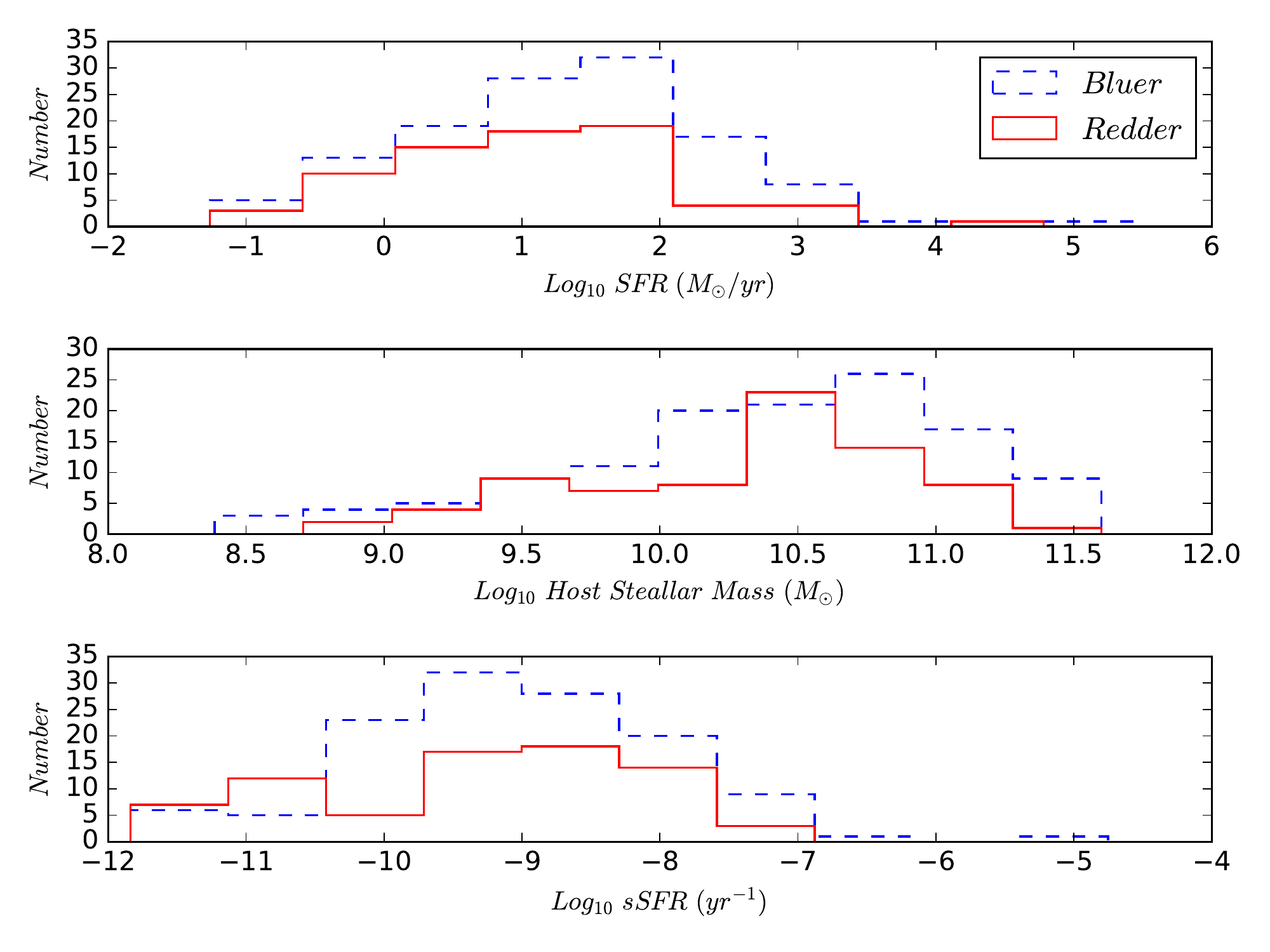}

    \caption{Left: SFR, stellar mass, and sSFR in faster and slower SN Ia hosts. In each case, two distributions seem to be different. See text for details. Right: Same as left but for bluer and redder SN Ia hosts. Here the distributions in each case seem not to differ.}
    \label{fig:sfrx1}
\end{figure*}

In Fig.~\ref{fig:sfrx1} we show distributions of SFR, stellar mass (from \citealt{uddin17}), and sSFR of host galaxies of faster, slower, bluer, and redder SNe Ia. Host galaxies of slower declining SNe Ia tend to have higher SFRs, lower masses, and therefore higher sSFRs. On the other hand, these host properties do not differ between bluer and redder SN Ia hosts.

We further elaborate these results by performing statistical analysis. First, we perform a Kolmogorov-Smirnov (K-S) hypothesis test to state if two host distributions for each host property are intrinsically the same or not. The null hypothesis in this case is that two distributions are drawn from the same parent distribution. We show K-S test results in Table \ref{ks}. 

It is evident from very low $p$-values for stellar mass and sSFR between faster and slower SN Ia hosts that they are intrinsically different. This also means that the mean values of stellar mass and sSFR on these two subgroups will be significantly different. We perform 100 bootstrap sampling to calculate mean and error in the mean  host property in each subgroup. They are shown in Table \ref{mean}. 

We see that on average slower declining SN Ia hosts have significantly higher ($5.4\sigma$) sSFR than faster declining SN Ia hosts. They also have significantly lower ($4.7\sigma$) stellar mass than their counterparts. These properties do not seem to differ between the hosts of bluer and redder SNe Ia.

\begin{table}
\begin{minipage}{\columnwidth}
\caption{K-S test results. Looking at the probabilities ($p$-values) it is clear that faster and slower SN Ia hosts differ significantly in terms of stellar mass and sSFR.}
\begin{center}
\begin{tabular}{clcc}
\hline
Subgroup Pairs & Host Property & $D$-statistics & $p$-value\\
\hline
Faster-Slower  & SFR &0.17 & 0.10\\
& Mass & 0.32 & 3e-5\\
&  sSFR &0.36 & 2e-6\\
\hline
Bluer-Redder & SFR & 0.17& 0.12 \\
& Mass & 0.19& 0.06\\
& sSFR &0.17 & 0.11\\
\hline

\end{tabular}
\end{center}
\label{ks}

\end{minipage}

\begin{minipage}{\columnwidth}
\caption{Mean and 1$\sigma$ error of host galaxy properties in various subgroups from 100 bootstrap sampling.}
\begin{center}
\begin{tabular}{llcc}
\hline
 Subgroup  & SFR & Mass & sSFR\\
 & $\rm (M_{\odot}/yr)$ & $\rm (M_{\odot})$ & $\rm (yr^{-1})$\\
\hline
Faster  & 0.986 (0.117) &10.581 (0.061) & -9.594 (0.116)\\
Slower  & 1.376 (0.102) &10.158 (0.068) & -8.773 (0.099)\\
Bluer & 1.305 (0.095) & 10.375 (0.061) & -9.064 (0.107)\\
Redder & 1.002 (0.135) & 10.308 (0.068) & -9.293 (0.140)\\
\hline

\end{tabular}
\end{center}
\label{mean}

\end{minipage}

\end{table}%

\section{Summary}\label{sum}
We have studied average spectra and physical properties of slower, faster, bluer, and redder SN Ia host galaxies. We have seen that the average spectrum of slower declining SN Ia hosts has stronger emission line features compared with the average spectrum of faster declining SN Ia hosts. They also have metallicities that are, on average, 0.24 dex lower than their counterparts as derived using pPXF. Between bluer and redder SN Ia hosts, average spectra do not vary in terms of emission line features. We also perform Lick index analysis of metal and Balmer lines. Our result show that hosts of faster declining SNe Ia have relatively higher metal content and have relatively older stellar population than hosts of faster declining SNe Ia. Similarly, redder SN Ia host have relatively more metal than bluer SN Ia hosts. We have calculated SFRs from $\rm H_{\alpha}$ luminosities and found that slower declining SN Ia hosts have significantly ($>5\sigma$) higher sSFR and also are significantly ($>4\sigma$) lower mass galaxies than faster declining SN Ia hosts. 

The K-S hypothesis test has shown that the hosts of slower and faster declining SNe Ia originate from different stellar populations when we compare their stellar mass and sSFR distributions. On the other hand, there are no significant differences in these properties between the hosts of bluer and redder SNe Ia and they originate from the same parent stellar population according to the K-S test.

Results we have summarized above have some significance in understanding SN Ia progenitors. It has been shown that young (prompt) SNe Ia originate from younger, low mass, actively star-forming galaxies (\citealt{childress14}). If we compare our results, we find that slower declining SNe Ia are young SNe Ia since they are hosted in high SFR galaxies.   

In this paper we have studied SN Ia light-curve properties. The next step should be to study average properties of SN Ia hosts in SN Ia corrected luminosity subgroups, which will be important for addressing systematic uncertainties in SN Ia cosmology. We have performed this study within a redshift of $z\sim 0.5$. The Dark Energy Survey (DES; \citealt{bernstein12}) will discover $\sim 3500$ SNe Ia upto redshift of $\sim 1.2$. The host spectra for these SN Ia are coming from OzDES (\citealt{yuan15}), which will enable us to continue this study at higher redshifts.

%%%%%%%%%%%%%%%%%%%%%%

\acknowledgments

\footnotesize{We thank Michael Childress and Chris Lidman  for helpful discussions. Syed A Uddin was supported by the Chinese Academy of Sciences President's International Fellowship Initiative Grant No. 2016PM014. This research made use of Astropy, a community-developed core Python package for Astronomy (\citealt{astropy13}). }

\bibliography{bibSyedUddin} % if your bibtex file is called example.bib

\end{document}